\documentclass[aps,prl,superscriptaddress,reprint,floats,citeautoscript,floatfix,nobibnotes,nofootinbib]{revtex4-1}

\usepackage[intlimits,tbtags]{amsmath}
\usepackage{times}
\usepackage{color}
\usepackage{amssymb}
\usepackage{graphicx}
\usepackage{bm}
\usepackage{array}
\usepackage{dcolumn}
\usepackage{bm}
\usepackage{setspace}

\newcolumntype{d}[1]{D{.}{.}{#1} }
\usepackage{xcolor}

\newcommand{\etal}{\emph {et al.}}

\newcommand{\jsnu}{Laboratory of Quantum Materials Design and Application, School of Physics and Electronic Engineering, Jiangsu Normal University, Xuzhou 221116, China}

\newcommand{\liaocheng}{Shandong Key Laboratory of Optical Communication Science and Technology, School of Physical Science \& Information Technology of Liaocheng University, Liaocheng 252059, China}

\begin{document}
\title{Seven-coordinated Silicon in a SiO$_2$He Compound Formed under the Extreme Conditions of Planetary Interiors}

\author{Shicong Ding}\affiliation{\jsnu}
\author{Pan Zhang}\affiliation{\jsnu}
\author{Kang Yang}\affiliation{\jsnu}
\author{Cailong Liu}\affiliation{\liaocheng}
\author{Jian Hao}\affiliation{\jsnu}
\author{Wenwen Cui}\email{wenwencui@jsnu.edu.cn}\affiliation{\jsnu}
\author{Jingming Shi}\email{jingmingshi@jsnu.edu.cn}\affiliation{\jsnu}
\author{Yinwei Li}\email{yinwei\_li@jsnu.edu.cn}\affiliation{\jsnu}\affiliation{\liaocheng}

\date{\today}

\begin{abstract}


Changes in atomic coordination numbers at high pressures are fundamental to condensed-matter physics because they initiate the emergence of unexpected structures and phenomena. Silicon is capable of forming eight-, nine-, and ten-coordinated structures under compression, in addition to the usual six-coordinated structures. The missing seven-coordinated silicon remains an open question, but here our theoretical study provides evidence for its existence at high pressures. A combination of a crystal-structure prediction method and first-principles calculations allowed prediction of a stable SiO$_2$He compound containing unique SiO$_7$ polyhedrons, which is a configuration unknown in any proposed silica phase. Consequently, seven-coordinated SiO$_7$ is a possible form of silica at high pressures. Further calculations indicate that the SiO$_2$He phase remains energetically stable with a solid character over a wide range of pressures exceeding 607 GPa and temperatures of 0–9000 K, covering the extreme conditions of the core–mantle boundary in super-Earth exoplanets, or even the Solar System’s ice giant planets. Our results may provide theoretical guidance for the discovery of other silicides at high pressures, promote the exploration of materials at planetary core–mantle boundaries, and enable planetary models to be refined.

\end{abstract}
\maketitle


Silica (SiO$_2$) is the main component of silicates in the core--mantle boundary (CMB) of the Earth and super-Earth exoplanets (1--10 times Earth's mass, M$_\oplus$). It has therefore attracted much recent attention in the fields of condensed matter physics, high-pressure physics, and planetary sciences~\cite{umemoto2006dissociation,tsuchiya2011prediction,millot2015shock}. Elucidating the high-pressure behavior of silica (like existence, form, and phase transition) is crucial for modeling the structure, formation, and evolution of planets. 
At ambient conditions, quartz is the most common stable phase of SiO$_2$~\cite{levien1980structure}. Compressing SiO$_2$ induces a complicated series of phase transitions; i.e., quartz phase $\rightarrow$ coesite phase ($\sim$2 GPa)~\cite{levien1981high} $\rightarrow$ stishovite phase ($\sim$10 GPa)~\cite{ross1990high,akaogi1984quartz} $\rightarrow$ CaCl$_2$-type phase ($\sim$48 GPa)~\cite{kingma1995transformation} $\rightarrow$ $\alpha$-PbO$_2$-type ($\sim$82 GPa)~\cite{driver2010quantum,dubrovinsky2001pressure} $\rightarrow$ pyrite-type phase ($\sim$260 GPa)~\cite{oganov2005structural,kuwayama2005pyrite}. 
These crystalline SiO$_2$ phases such as CaCl$_2$-, $\alpha$-PbO$_2$-, and pyrite-type have a same Si coordination number of six while for the glassy state of SiO$_2$ may structurally evolve above 95 GPa to have a Si coordination number greater than six~\cite{kono2020structural} which is proposed by experiment.
Recent work has predicted a silica $R\bar{3}$ phase with mixed six-, eight-, and nine-fold coordination stabilized at 645 GPa, which subsequently transforms into the Fe$_2$P-type structure at 890 GPa~\cite{liu2021mixed}. With respect to temperature, the cotunnite-type phase is much more stable than the $R\bar{3}$ or Fe$_2$P-type phases of SiO$_2$~\cite{umemoto2006dissociation,tsuchiya2011prediction,wu2011identification}. Coordination in the cotunnite- and Fe$_2$P-type phases can reach nine under extremely high pressure and temperature. Finally, the Fe$_2$P-type structure transforms into the $I$4/$mmm$ phase with a coordination number of 10 at 10 TPa~\cite{lyle2015prediction}. 

Planetary interiors are naturally extremely hot, pressurized, and rich in silica. Understanding the silica structure before it melted under extreme conditions is the key to determining a planet's internal structure and evolution. In 2015, Millot \etal's laser-driven shock experiment found that the melting temperature of silica rises to 8300 K at a pressure of 500 GPa, comparable to the CMB of a super-Earth of 5M$_\oplus$~\cite{millot2015shock}. However, laboratory experiments face several challenges when reproducing the conditions of giant planets or super-Earths~\cite{duffy20152}. 
Theoretical modeling can effectively simulate extreme conditions and so aid exploration of the behavior of minerals such as SiO$_2$ inside planetary interiors~\cite{scipioni2017electrical,gonzalez2016melting,niu2015prediction,tsuchiya2011prediction,wu2011identification}. 
Using first-principles molecular dynamics simulations, Scipionia \etal~found that under conditions of the early Earth's and super-Earths' deep molten mantles, the electron conductivity of liquid SiO$_2$was sufficiently large to support a silicate dynamo.~\cite{scipioni2017electrical}. 
Felipe \etal~studied the melting curves of SiO$_2$ up to 6 TPa using first-principles molecular dynamics simulations, finding that SiO$_2$ is solid under the conditions of the cores of all the solar system's gas giants and also super-Earths of 10 M$_\oplus$~\cite{gonzalez2016melting}. Taku \etal~and Wu \etal~independently proposed that the cores of Uranus and Neptune are expected to consist of Fe$_2$P-type silica and MgSiO$_3$~\cite{tsuchiya2011prediction,wu2011identification}. Niu \etal's structural prediction of the Mg--Si--O ternary phase diagram under exoplanet pressure revealed two novel compounds, Mg$_2$SiO$_4$ and MgSi$_2$O$_5$, as being stable at the CMB of large super-Earths~\cite{niu2015prediction}. Overall, these studies make important contributions to the understanding of the physicochemical diversity of planetary systems.

Helium is the second most abundant element in the universe, and is extensively present in stars and gas giant planets~\cite{guillot2004probing,stevenson2008metallic}. It is usually chemically inert and does not react easily with other substances at ambient conditions due to its closed-shell electronic configuration. The melting and volcanic effects of the early Earth's mantle caused He in the upper mantle to escape via internal degassing and disappear into space, while a layer of the lower mantle continues to retain He~\cite{jackson2017primordial}. Recent studies have shown that FeO$_2$ can react with He near the Earth's CMB to trap He in the lower mantle~\cite{zhang2018rare}. Ice giants are mainly composed of H$_2$O, NH$_3$, and CH$_4$; their atmospheres are rich in He~\cite{guillot2004probing,nettelmann2016uranus}. Theoretical studies have shown that H$_2$O~\cite{liu2019multiple,liu2015stable,bai2019electrostatic}, NH$_3$~\cite{shi2020formation,bai2019electrostatic,liu2020plastic}, and CH$_4$~\cite{gao2020coexistence} can form stable compounds with He at high pressure, and first-principles molecular dynamics suggests that both H$_2$O--He and NH$_3$--He may exist in a superionic state in the interiors of icy planets, in addition to the plastic state of NH$_3$--He and the plastic or partially diffusive state of CH$_4$--He at lower pressures and temperatures. 
Theoritical study reported that He can form stable crystalline compounds with iron at terapascal pressures~\cite{monserrat2018helium}.
Other recent studies have reported the high-pressure stability of He--alkali oxides (sulfides)~\cite{gao2019prediction}, HeN$_4$~\cite{li2018route}, Na$_2$He~\cite{dong2017stable}, Mg(Ca)F$_2$He~\cite{liu2018reactivity}, and XeHe$_2$~\cite{wang2015prediction}.

\begin{figure}[htp]
\centering
  \includegraphics[width=0.99\linewidth,angle=0]{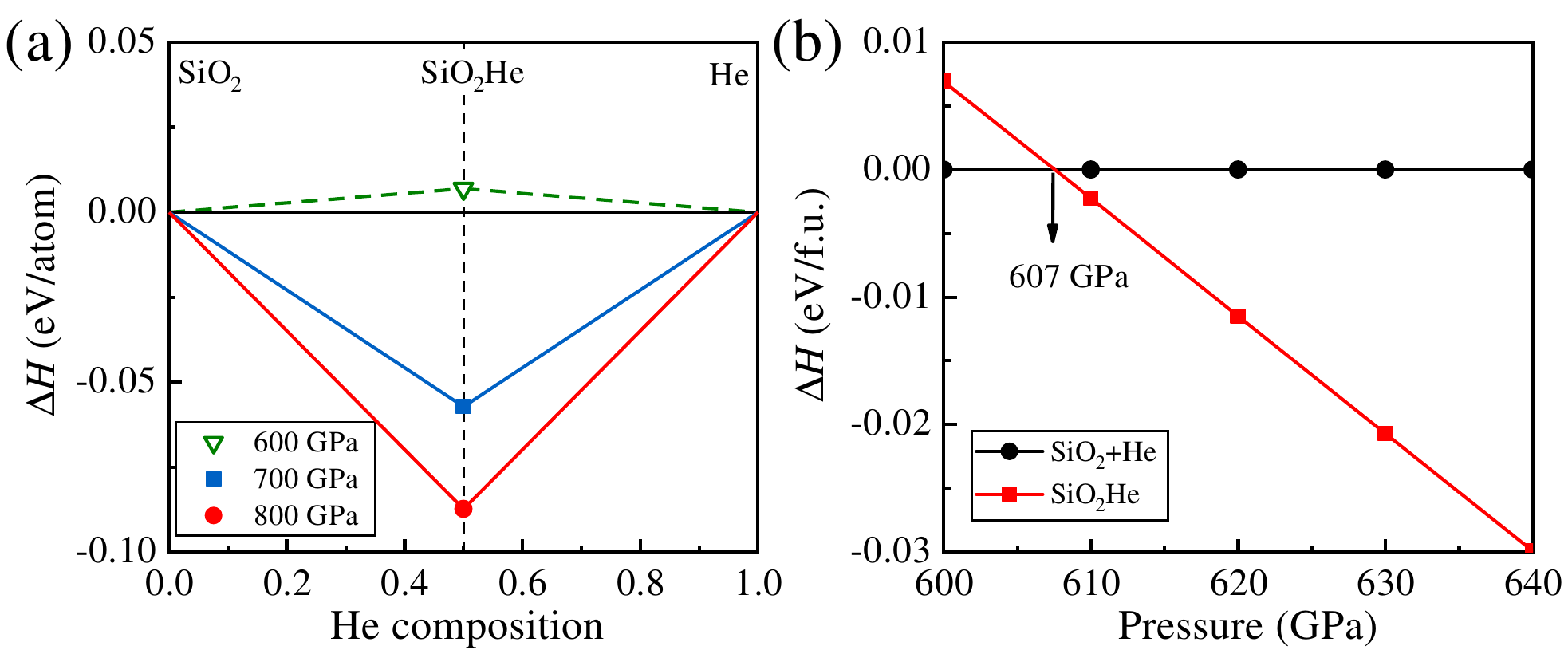}
  \caption{\label{fig:1} \textbf{(a)} Convex hull for formation enthalpies ($\Delta H$) of the SiO$_2$--He system at selected pressures, defined as $\Delta H$ = [$H$((SiO$_2$)$_xHe_y$) -- $xH$(SiO$_2$) -- $yH$(He)] / (3$x$ + $y$). \textbf{(b)} Calculated formation enthalpy of SiO$_2$He relative to decomposition products SiO$_2$ (pyrite-type at 600 GPa, $R\bar{3}$ at 700 and 800 GPa) and He ($hcp$at 600--800 GPa) as a function of pressure.}
\end{figure}

Previous studies have shown that the insertion of He into cristobalite can greatly decrease its compressibility, resulting in two cristobalite--helium compound phases; i.e., I ($P$2$_1$/$c$, stable above 6.4 GPa) and II ($R$$\bar{3}$$c$, at 1.7--6.4 GPa)~\cite{sato2013anomalous,matsui2014crystal}. Despite substantial experimental and theoretical efforts to understand the compositions of planetary interiors, the results are far from sufficient. The nature of SiO$_2$ under extreme conditions, especially in combination with other elements, in planetary interiors requires further investigation; its elucidation would greatly aid our understanding of planetary formation and evolution. 

Combining density function theory and structural prediction, this work performs an extensive exploration of the SiO$_2$--He systems at high pressure and high temperature. The results show that He can react with SiO$_2$ to form $Pnma$ SiO$_2$He at high pressures, and first-principles molecular dynamics simulations indicate that solid $Pnma$ SiO$_2$He exists in the mantle of Uranus and Neptune, as well as the CMB of Saturn and super-Earth exoplanets. Interestingly, $Pnma$ SiO$_2$He has four equivalent seven-coordinated SiO$_7$ octahedra. Previously, only six-, eight-, nine-, and 10-coordination has been reported in the stable high-pressure phases of SiO$_2$, with seven-coordinated SiO$_2$ as a metastable phase~\cite{tsuchiya2011prediction,liu2021mixed}. Our study suggests that seven-coordinated SiO$_7$ might be present in $Pnma$ SiO$_2$He at high pressure.
  
\begin{figure}[htp]
\centering
  \includegraphics[width=0.90\linewidth,angle=0]{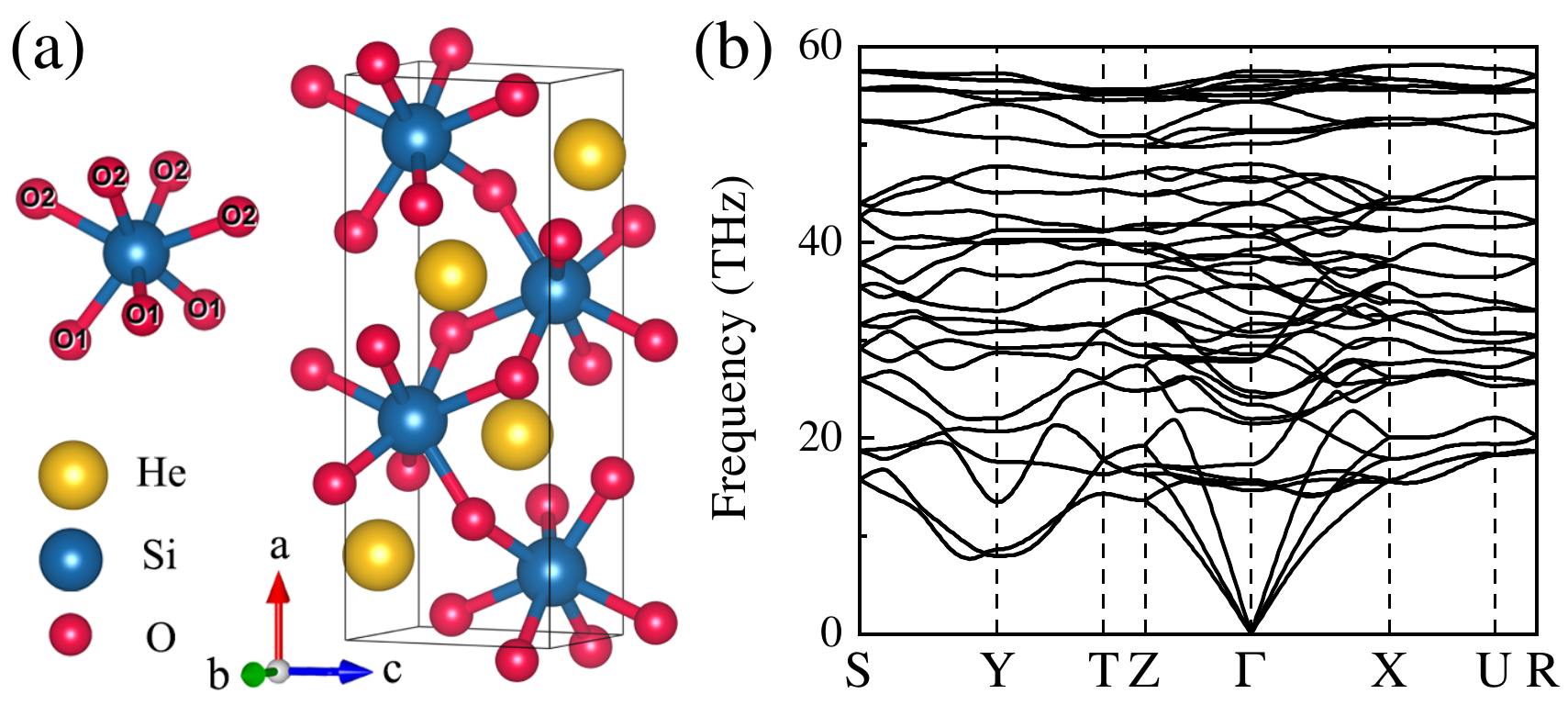}
  \caption{\label{fig:2} \textbf{(a)} Crystal structure predicted for $Pnma$ SiO$_2$He at 700 GPa and the equivalent SiO$_7$ octahedron. Yellow, blue, and red spheres represent He, Si, and O atoms, respectively. \textbf{(b)} Calculated phonon dispersions of $Pnma$ SiO$_2$He at 700 GPa.}
\end{figure}

Structure predictions for the SiO$_2$--He system are performed here using CALYPSO~\cite{wang2010crystal,wang2012calypso}, based on a global minimization of free energy surfaces in conjunction with $ab~initio$ total-energy calculations, which has been used successfully to predict various systems under high pressure~\cite{li2014metallization,li2015high,li2016dissociation,cui2019role}. Structural relaxations are performed using density functional theory as implemented in the Vienna $ab~initio$ simulation package~\cite{kresse1996efficient,kresse1996efficiency}, adopting the Perdew--Burke--Ernzerhof exchange--correlation functional under the generalized gradient approximation~\cite{perdew1992atoms,perdew1996generalized}. All-electron projector augmented wave pseudopotentials with 3$s^2$3$p^2$, 2$s^2$2$p^4$, and 1$s^2$ valence configurations are chosen for Si, O, and He atoms, respectively~\cite{kresse1999ultrasoft}. Structure searches use a plane wave cutoff energy of 1000 eV and k-point mesh of 2$\pi$ $\times$ 0.04 \AA$^{-1}$. The lowest-energy structural optimization and electronic properties are calculated using `hard' pseudopotentials with 2$s^2$2$p^6$3$s^2$3$p^2$ and 2$s^2$2$p^4$ valence configurations for Si and O atoms, respectively. A plane wave cutoff energy of 1500 eV and k-point mesh of 2$\pi$ $\times$ 0.03 \AA$^{-1}$ are set to ensure total energy and forces convergence better than 1 meV/atom and 1 meV/\AA, respectively. Phonon calculations are carried out using a supercell approach, as implemented in PHONOPY code~\cite{togo2008first}. $Ab~initio$ molecular dynamics (AIMD) simulations~\cite{nose1984unified} explore the form of SiO$_2$He compounds at high pressures and high temperatures. A 2 $\times$ 2 $\times$ 2 supercell of 288 atoms is employed, along with $\Gamma$ point sampling of the Brillouin zone. The canonical $NVT$ ($N$-number of particles, $V$-volume, and $T$-temperature) ensemble uses a Nose--Hoover thermostat~\cite{hoover1985canonical} with SMASS = 2, and each simulation consists of 20,000 time steps of 0.5 fs. We allow 1 ps for thermalization, and then exact data for 4 ps. The crystal structures and electron localization function (ELF) are plotted using VESTA software~\cite{momma2011vesta}.

Extensive structure searches for (SiO$_2$)$_x$He$_y$ ($x$, $y$ = 1--3) are performed at 100--800 GPa with maximum simulation cells up to four formula units (f.u.) at each composition. Fig.~\ref{fig:1}(a) summarizes the formation enthalpies ($\Delta H$) of the stoichiometries with respect to decomposition into SiO$_2$ and He, with positive formation enthalpies above 0.05 eV/atom not shown in the convex hull. The stoichiometry of SiO$_2$He is identified with a negative formation enthalpy at 700 and 800 GPa, whose $\Delta H$ approaches zero at $\sim$607 GPa, as shown in Fig.~\ref{fig:1}(b). This reveals that the SiO$_2$He compound is thermodynamically stable when the pressure is above the critical pressure. 

\begin{figure}[htp]
\centering
  \includegraphics[width=1.0\linewidth,angle=0]{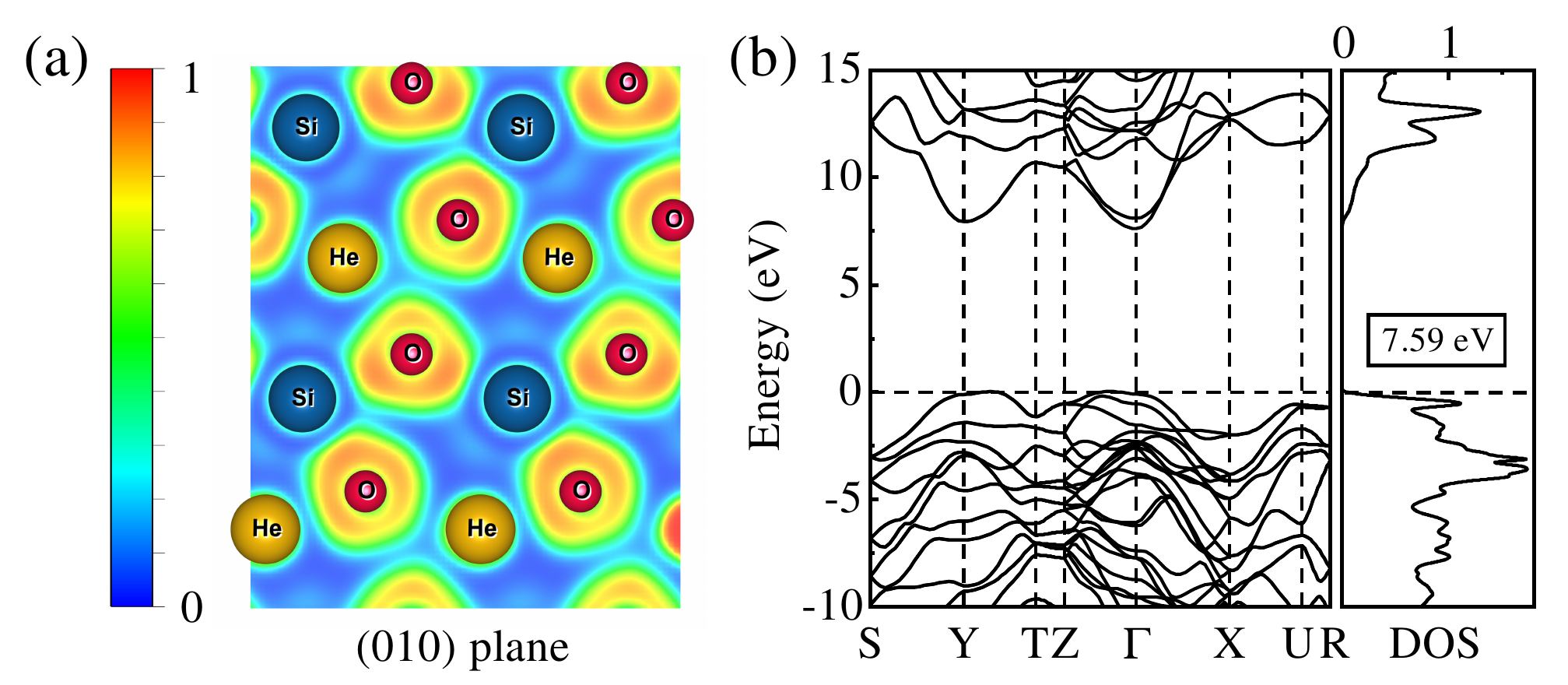}
  \caption{\label{fig:3} \textbf{(a)} ELF and \textbf{(b)} electronic band structures (left) and density of states (DOS; right) of $Pnma$ SiO$_2$He at 610 GPa.}
\end{figure}

The predicted SiO$_2$He phase has an orthorhombic structure with space group $Pnma$ (4 f.u. in a unit cell) and remains stable up to 1000 GPa, the maximum pressure studied here.
Fig.~\ref{fig:2}(a) shows the structural configuration of $Pnma$ SiO$_2$He, and Table S1 in the Supplemental Material summarizes the detailed structural parameters~\cite{supple}. 
By further analysis of the atomic positions, we find that all the silicon atoms in our predicted $Pnma$ SiO$_2$He phase are with seven coordination. 
Each Si atom and its seven surrounding O atoms form a SiO$_7$ polyhedron, which is never reported by previous studies in Si-O systems and reveals the emergence of a novel mode of bonding for Si--O  under high pressure.
 Four SiO$_7$ polyhedra in the unit cell are equivalent with Si--O bond lengths ranging from 1.54 to 1.67 \AA. Two SiO$_7$ polyhedra in the center of the cell connect with each other by sharing edges and connect to the top and bottom sides by sharing vertices. There are two kinds of O atom, which are located in the 4$c$ Wyckoff positions. 
Surprisingly, the He atoms occupy positions similar to Si atoms in the SiO$_7$ polyhedra. If considering only the relative positions of He and O atoms, they can also form a HeO$_7$ polyhedral structure nearly identical to the SiO$_7$ polyhedra with He--O distances ranging from 1.51 to 1.71 \AA. The volumes of the polyhedra are 5.85 \AA $^3$ for HeO$_7$ and 5.79 \AA $^3$ for SiO$_7$ (see Figure S1 in the Supplemental Material~\cite{supple}). If we take the $Pnma$ SiO$_2$He phase as prototype structure and choose Si atoms to substitute all the He atoms, we can obtain a novel  SiO compounds with the same $Pnma$ symmetry. Unfortunately, calculation results show that the formation enthalpy of this new SiO compound is about 0.71 eV/atom with respect to SiO$_2$ and Si, while about -0.76 eV/atom with respect to elements at 700 GPa, which reveals that the designed structure might be a metastable phase of Si-O compound. If this phase can be synthesized on experiment, which reveals that a novel unconventional silicon coordination Si-O compound is exist under high pressure.  
Phonon calculations reveal that there are no imaginary dispersions of $Pnma$ SiO$_2$He at 700 (Fig.~\ref{fig:2}(b)), 610, or 800 GPa (Supplemental Material Figure S2~\cite{supple} for both), confirming the robust dynamic stability of $Pnma$ SiO$_2$He.

Fig.~\ref{fig:3}\textbf{(a)} shows the ELF of $Pnma$ SiO$_2$He to analyze the bonding features of He inserted in SiO$_2$. The electrons are strongly localized in O atoms and delocalized around Si, indicating ionic bonding between these atoms, which is consistent with our Bader charge analysis calculation results~\cite{henkelman2006fast}. The Bader charge transfer calculations for the SiO$_2$He structure at 610 GPa show that a Si atom loses $\sim$3.39 $e^-$, 1.64 $e^-$ of which is transferred to oxygen atom O$_1$, 1.67 $e^-$ to oxygen atom O$_2$, and only $\sim$0.08 $e^-$ to He. The near complete absence of charge transfer between SiO$_2$ and He indicates that He atoms do not chemically bond with neighboring atoms.
Other He-containing systems behave similarly: for example, FeO$_2$He~\cite{zhang2018rare}, H$_2$O--He~\cite{liu2019multiple,liu2015stable,bai2019electrostatic}, NH$_3$--He~\cite{bai2019electrostatic,shi2020formation,liu2020plastic}, HeN$_4$~\cite{li2018route}, and Na$_2$He~\cite{dong2017stable}. Electronic band structure and DOS calculations
show that $Pnma$ SiO$_2$He is an insulator with a band gap of 7.59 eV at 610 GPa (Fig.~\ref{fig:3}\textbf{(b)}). Further compression slightly increases the band gap to 8.26 eV at 1 TPa (cf. Figure S3 in the Supplemental Material~\cite{supple}).

\begin{figure}[htp]
\centering
  \includegraphics[width=0.80\linewidth,angle=0]{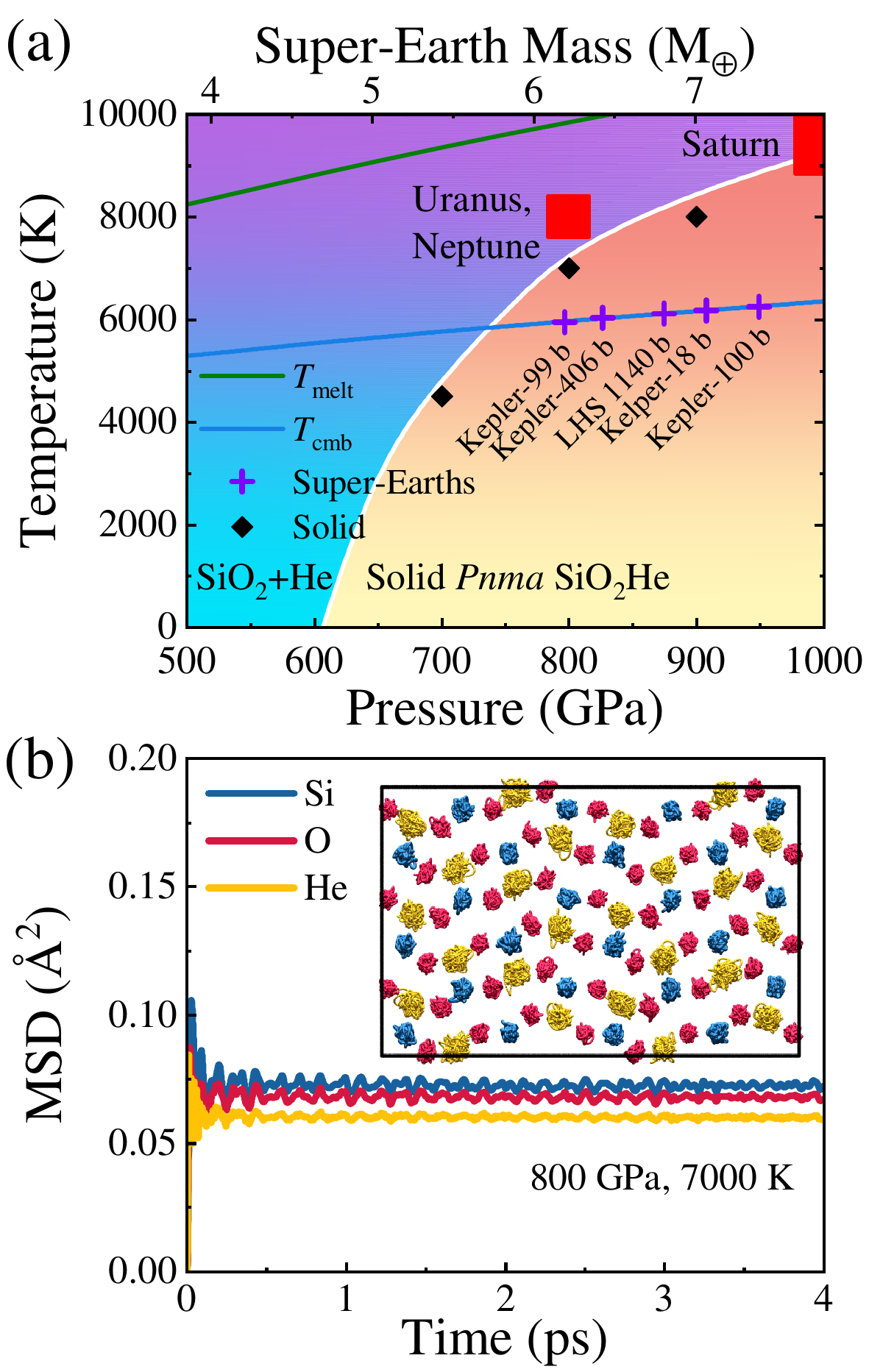}
  \caption{\label{fig:4} \textbf{(a)} Pressure--temperature ($P$--$T$) phase diagram showing the stability field of $Pnma$ SiO$_2$He and its boundary (white line) with the decomposition products SiO$_2$ + He. Black diamonds represent solid $Pnma$ SiO$_2$He. Red squares indicate the CMB of Uranus, Neptune, and Saturn~\cite{guillot1999interiors}. The proposed silica melting line ($T_m = 1968.5 + 307.8P_m^{0.485}$)~\cite{millot2015shock} and the super-Earths’ mantle adiabat ($T_{cmb} =2275.2 + 1583.5 \left(\frac{P}{129.8}\right) ^{0.485}$)~\cite{wagner2012rocky} are plotted in green  and blue, respectively. $P = 129.8 \left(\frac{M_{S}}{M_E}\right)$ links the pressure scale to the super-Earth mass ($M_S$) scale~\cite{wagner2012rocky}. Purple crosses indicate some Earth-like super-Earths with pressure corresponding to that in $T_{cmb}$. \textbf{(b)} Calculated mean squared displacement (MSD) of the atomic positions of $Pnma$ SiO$_2$He at 800 GPa and 7000 K. Insert gives atomic trajectories in one supercell from the simulations in the last 2 ps.}
\end{figure}

The interiors of planets harbor conditions of high pressure and high temperature.
In order to determine the stable regions of pressure and temperature for $Pnma$ SiO$_2$He, we consider the temperature effect  using the harmonic approximation by performing the phonon dispersion calculations for all the structures in 600, 700, 800, 900 and 1000 GPa.  We calculate the formation energy of $Pnma$ SiO$_2$He with respect to the decomposition products SiO$_2$ and He under high pressure and temperature.The critical boundary of stability (white line) is plotted in the phase diagram of the SiO$_2$ system, as shown in Fig.~\ref{fig:4}\textbf{(a)}. For example, our calculations show that the $Pnma$ SiO$_2$He phase is stable at 800 GPa when the temperature is below 7000 K, above that it will decompose into SiO$_2$ and He.
AIMD simulations consider the form of $Pnma$ SiO$_2$He at the conditions of giant planets; i.e., pressures of 600--1000 GPa and temperatures of 0--10,000 K. Fig.~\ref{fig:4}\textbf{(b)} shows the calculated mean squared displacement (MSD) of atomic positions and atoms trajectories for Si (blue), O (red), and He (orange). At $P$ = 800 GPa and $T$ = 7000 K, 
all atoms in SiO$_2$He vibrate around their lattice positions and the diffusion coefficients are zero ($D^{Si} = D^O = D^{He} = 0$), indicating that the $Pnma$ SiO$_2$He phase remains solid at these conditions.
Fig.~\ref{fig:4}\textbf{(a)} can clearly depict its form inside gas giants and super-Earths. The yellow gradient area in the figure indicates $Pnma$ SiO$_2$He as being a stable solid. 
As the super-Earths' mantle adiabat crosses the region of stable $Pnma$ SiO$_2$He, $Pnma$ SiO$_2$He may be present within the CMBs of super-Earths larger than 5.7 M$_\oplus$. Potential examples include exoplanets Kepler-99 b, Kepler-406 b, LHS 1140 b, Kepler-18 b, and Kepler-100 b (purple crosses in the figure). 
Figure S4 in the Supplemental Material~\cite{supple} shows mass--radius relations for these Earth-like structures. 
Recent studies have indicated an unclear distinction between the mantle and the core of Uranus and Neptune~\cite{helled2020uranus}; the CMB of both is slightly above the region of $Pnma$ SiO$_2$He, suggesting that $Pnma$ SiO$_2$He may be present in both planets' mantles. 
Saturn is a gas giant with much He in its mantle. Some of its CBM overlaps with the stable region of $Pnma$ SiO$_2$He, suggesting that $Pnma$ SiO$_2$He may also be present in Saturn's CBM. 
We also calculate the MSD of $Pnma$ SiO$_2$He at high temperatures to study each atom's behavior, as shown in Fig. S5 and Fig. S6. 
The calculations show that $P$ = 700 GPa, $T$ = 4500 K and $P$ = 900 GPa, $T$ = 8000 K can be viewed as critical conditions of $Pnma$ SiO$_2$He transitioning from a solid to a superionic phase.
At 800 GPa, $Pnma$ SiO$_2$He remains solid up to 8000 K, and heating to 9000 K transforms it to a superionic phase with partial He diffusion (Figure S6(b)~\cite{supple}) with a diffusion coefficient of $D^{He}$ = 1.31 $\times10^{-6}$ cm$^2$s$^{-1}$. Further heating to 10,000 K increases the diffusion of He; its coefficients becomes $D^{He} = 3.43 \times10^{-6}$ cm$^2$s$^{-1}$. At 12,000 K, all atoms are diffusive (see Fig. S6(d)), with higher diffusion coefficients of $D^{Si} =8.91 \times10^{-5}$ cm$^2s^{-1}$, $D^{O} =1.07 \times10^{-4}$ cm$^2s^{-1}$, and $D^{He} =1.44 \times10^{-4}$ cm$^2s^{-1}$.
Previous studies show that with the  increasing temperature, the H atoms in CH$_4$-He~\cite{gao2020coexistence} and H$_2$O-He~\cite{liu2019multiple,liu2015stable,bai2019electrostatic} systems become diffuse firstly and transform into superionic phase, while the He atoms are easier to diffuse in NH$_3$-He~\cite{shi2020formation,bai2019electrostatic,liu2020plastic}. 
As compared to these He-incorporation compounds, the critical temperatures transition from a solid to a superionic or fluid phase for SiO$_2$He are higher than them, displaying the He atoms diffusing firstly, which is mainly caused by the following two reasons. First, the  atomic mass of He atom  is larger than that of H atom. Second, the HeO$_7$ polyhedron configuration in the SiO$_2$He unit cell makes the He atoms be trapped inside the cage, which restricts its diffusion.
Figure S7 shows the estimated pressure--temperature conditions for the solid, superionic, and fluid phases of SiO$_2$He and reveals that the incorporation of He notably inhibits the melting of silica.

In conclusion, combining first-principles theory and structural prediction, we report a $Pnma$ SiO$_2$--He phase that is stable under planetary interior conditions. This phase is stable over a large pressure range from 607 GPa to at least 1 TPa at 0 K. It is composed of four equivalent seven-coordinated SiO$_7$ polyhedra, which have not been found previously in the ground state of silica. AIMD calculations show the possibility of solid SiO$_2$He being present at the CMBs of the gas planets of our solar system as well as in super-Earths more massive than 5.7 M$_\oplus$. The present results not only benefit our understanding of the high-pressure behavior of materials, but can also broaden the family of He-containing compounds and provide assistance in understanding the structural models and evolution of giant planets' interiors.


The authors acknowledge funding from the NSFC under grants No. 12074154, No. 11804129, No. 11722433 and No. 11804128, and the funding from the Science and Technology Project of Xuzhou under grant No. KC19010. Y.L. acknowledges the funding from the Six Talent Peaks Project and 333 High-level Talents Project of Jiangsu Province. S.D. acknowledges the founding from Postgraduate Research \& Practice Innovation Program of Jiangsu Province No. KYCX20\_2223. All the calculations were performed at the High Performance Computing Center of the School of Physics and Electronic Engineering of Jiangsu Normal University.







\bibliography{paper}

\end{document}